\documentclass[journal,10pt,comsoc]{IEEEtran}
\usepackage[T1]{fontenc}
\usepackage{psfrag,amsmath,amssymb,color,cite, amsmath,graphicx}
\usepackage{amsthm}
\usepackage{listings}
\usepackage{stfloats}
\usepackage{mathtools}

\newcommand{\DeclareAutoPairedDelimiter}[3]{%
  \expandafter\DeclarePairedDelimiter\csname Auto\string#1\endcsname{#2}{#3}%
  \begingroup\edef\x{\endgroup
    \noexpand\DeclareRobustCommand{\noexpand#1}{%
      \expandafter\noexpand\csname Auto\string#1\endcsname*}}%
  \x}
\DeclareAutoPairedDelimiter\modulo{[}{]} 
\usepackage{fancyhdr,graphicx,amsmath,amssymb}
\usepackage[linesnumbered,ruled,vlined]{algorithm2e}
\include{pythonlisting}
\usepackage{mathrsfs}
\usepackage{multirow}
\usepackage{pifont}
\usepackage[colorinlistoftodos]{todonotes}

\newcommand{\beqn}{\begin{equation}}
\newcommand{\eeqn}{\end{equation}}
\newcommand{\beqa}{\begin{eqnarray}}
\newcommand{\eeqa}{\end{eqnarray}}
\newcommand{\beqas}{\begin{eqnarray*}}
\newcommand{\eeqas}{\end{eqnarray*}}

\presetkeys{todonotes}{color=green!30}{}
\usepackage{amsmath}
\usepackage{amsfonts}
\usepackage{graphicx}
\usepackage{array}
\usepackage{ amssymb }
\newcolumntype{P}[1]{>{\centering\arraybackslash}p{#1}}
\newcolumntype{M}[1]{>{\centering\arraybackslash}m{#1}}

\begin{document}

\title{Low-Complexity Linear Diversity-Combining Detector for MIMO-OTFS}
\author{\IEEEauthorblockN{Tharaj Thaj and Emanuele Viterbo
\\}
\IEEEauthorblockA{ECSE Department, Monash University, Clayton, VIC 3800, Australia\\
Email: \{tharaj.thaj, emanuele.viterbo\}@monash.edu}}

\maketitle
\begin{abstract}
This paper presents a low complexity detector for multiple-input multiple-output (MIMO) systems based on the recently proposed orthogonal time frequency space (OTFS) modulation. In the proposed detector, the copies of the transmitted symbol-vectors received through the different diversity branches (propagation paths and receive antennas) are linearly combined using the maximum ratio combining (MRC) technique to iteratively improve the signal to interference plus noise ratio (SINR) at the output of the combiner. {\color{black} To alleviate the performance degradation due to spatial correlation 
at the receiver antennas, we present a sample-based method to estimate such correlation and find the optimized combining weights for MRC from the estimated correlation matrix.} The detector performance and complexity improve over the linear minimum mean square error (LMMSE) and message passing (MP) detectors proposed in the literature for MIMO-OTFS. 
\end{abstract}
\vspace{-3mm}
\section{Introduction}\footnote{This work was supported by the Australian Research Council through
the Discovery Project under Grant DP200100096}
Reliable communications in high mobility wireless channel scenarios are essential to satisfy the wireless network requirements of 6G mobile communications. 
Orthogonal frequency division multiplexing (OFDM), at the core of the physical layer of 4G/5G, multiplex information symbols on closely spaced orthogonal sub-carriers. This results in performance degradation in high-mobility wireless channels due to the loss of orthogonality among the sub-carriers.

The recently proposed orthogonal time frequency space (OTFS) modulation is based on the idea of multiplexing the information symbols in the delay-Doppler (DD) domain, resulting in significant performance gains over OFDM in high-mobility channels, \cite{Hadani}. In OFDM, each information symbol is transmitted over a single time-frequency resource, which is susceptible to frequency and time-selective fading effects, whereas OTFS  multiplexes each information symbol over 2D orthogonal basis functions that span the entire time and frequency domain, but are localized in the DD domain. As a result, all information symbols experience a constant flat fading equivalent channel.

Multiple-input multiple-output (MIMO) based on OTFS (MIMO-OTFS) can further increase the spectral efficiency to meet the data rate demands of 6G. The superior performance of MIMO-OTFS over MIMO-OFDM and its system model, detection, and channel estimation methods have been well studied in the literature, \cite{Choks_MIMO1,Choks_MIMO2,Choks_MIMO3,Farhang,  Saif,Prem}. The biggest challenge with a MIMO system is the high processing complexity at the receiver. 
 The message passing (MP) algorithm proposed for MIMO-OTFS in \cite{Choks_MIMO1} offers excellent performance but still suffers from high complexity, especially in high Doppler spread channels and high-order modulation. Further, spatial correlation can degrade the error performance of the MIMO system, especially if the correlation of the antennas is not taken into account while designing receiver algorithms, \cite{Lokya01a}.

{\color{black} In this letter, we propose a linear-complexity detector for MIMO-OTFS with {\color{black} rectangular pulse-shaping waveform}, based on the maximum ratio combining (MRC) diversity-combining principle proposed for SISO-OTFS in \cite{Rake_DFE}, to efficiently combine the distinct antenna and multipath copies of the transmitted symbols.  Differently from \cite{Rake_DFE}, the combining weights of the MRC algorithm are optimized 
to combat the adverse effects of spatial correlation at the receiver (Rx). We further propose a sample-based method to estimate the spatial correlation between the channels from the estimated channel coefficients at different Rx antennas. Further, we analyze via simulations the performance of the proposed detection method with spatial correlation and practical channel estimation at the Rx. Finally, we show that the proposed detector is linear in the number of transmitted symbols and antennas, making the detector complexity convenient even for large MIMO systems.  
}
\vspace{-0.5mm}
{\bf Notations}: The following notations will be used: $a$, $\bf{a}$, ${\bf A}$ represent a scalar, vector, and matrix, respectively; ${\bf a}[n]$ and ${\bf A}[m,n]$ represent the $n$-th and $(m,n)$-th element of ${\bf a}$ and ${\bf A}$, respectively; ${\bf A}^\dag$, ${\bf A}^*$ and ${\bf A}^n$ represent the Hermitian transpose, complex conjugate and $n$-th power of ${\bf A}$. The set of $M \times N$ dimensional matrices with complex entries is denoted by ${\mathbb{C}}^{M \times N}$. Let {\color{black}$\otimes$ denote the Kronecker product}, $|\mathcal{S}|$ the cardinality of the set $\mathcal{S}$, and vec$({\bf A})$ the column-wise vectorization of the matrix ${\bf A}$. Let ${\bf F}_N$ be the {\color{black} normalized} $N$ point discrete Fourier transform (DFT) matrix and ${\bf I}_M$ the $M \times M$ identity matrix.

\vspace{-2mm}
\section{System Model}
Consider a MIMO-OTFS system with $n_{\rm T}$ and $n_{\rm R}$ transmit and receiver antennas, respectively. Let ${\bf X}^{(t)}$ and ${\bf Y}^{(r)}$ be the  $M \times N$ DD domain OTFS information symbols, transmitted from the $t$-th antenna and received at the $r$-th antenna, respectively. All transmitted frames of duration $NT\,$[s] occupy the same bandwidth of $M \Delta f\,$[Hz], with $T\Delta\! f=1$. The time-domain samples transmitted from the $t$-th antenna and the received samples at the $r$-th antenna are given by
\begin{equation}
    {\bf s}^{(t)}={\rm vec}({\bf X}^{(t)}\cdot{\bf F}_N^{\dag}),\quad{\bf r}^{(r)}={\rm vec}({\bf Y}^{(t)}\cdot{\bf F}_N^{\dag})
    \label{MIMO_vec}
\end{equation}
At the transmitter, a cyclic prefix (CP) or zero-padding (ZP) of length $L_{\rm G}$ greater than the channel delay spread is inserted before each of the $N$ time-domain blocks of the OTFS frame. 
\subsection{Channel}
Let $h_i^{(r,t)}$, $\tau_i^{(r,t)}$ and $\nu_i^{(r,t)}$ be the complex path gain, delay and Doppler shift, respectively, associated with the $i$-th path in the channel between the $r$-th receive antenna and the $t$-th transmit antenna (the $(r,t)$ sub-channel). The DD domain representation of the MIMO multipath channel is given  by
\begin{equation}
    h^{(r,t)}(\tau,\nu)=\sum_{i=1}^{P^{(r,t)}}h_i^{(r,t)}\delta(\tau-\tau_i^{(r,t)})\delta(\nu-\nu_i^{(r,t)})
\label{eq:DD_io_hrt}
\end{equation}
for $t=1,\ldots, n_{\rm T}$ and $r=1,\ldots, n_{\rm R}$, 
where $P^{(r,t)}$ is the number of DD domain paths in the $(r,t)$ sub-channel. The corresponding delay-time (DT) channel can be written as
\begin{align}
    g^{(r,t)}(\tau,\theta)= \int_\nu h^{(r,t)}(\tau,\nu){\rm e}^{j2\pi\nu (\theta-\tau)}\, d\nu \label{eq:cont-time}
\end{align}
where $\theta$ is the continuous-time variable.

The receiver samples the incoming signals at integer multiples of the sampling period $1/M \Delta\! f$. Let $\ell_i={\tau_i}{M\Delta f}$ and $\kappa_i={\nu_i}{NT}$ be the {\em normalized delay} and {\em normalized Doppler-shift} associated with the $i$-th path. Following \cite{Rake_DFE}, the discrete-time equivalent channel is obtained by sampling (\ref{eq:cont-time}) at times $\theta=\frac{q}{M\Delta f}$ and delays $\tau=\frac{\ell}{M\Delta f}$ with $q,\ell \in \mathbb{Z}$ as
\begin{align}
    {g}^{(r,t)}[\ell,q]&= \sum_{i=1}^{P^{(r,t)}}h_i^{(r,t)}z^{(q-\ell)\kappa_i^{(r,t)}}{\delta}[\ell-\ell_i^{(r,t)}]   \label{MIMO_DT_eqn_int}
\end{align}
where we assume integer {\em normalized delays}, i.e., $\ell_i^{(r,t)} \in \mathbb{Z}$ and $z={\rm e}^{\frac{j2\pi}{MN}}$. No assumption is made on $\kappa_i^{(r,t)}$ to be integer.
\vspace{-3mm}
{\color{black}
\subsection{Spatial correlation}
In MIMO systems, there is often correlation among the $(r,t)$ sub-channels depending on the propagation environment, antenna patterns, and the relative locations of the Tx and Rx antennas. If we assume that the transmitter and receiver are sufficiently separated, then the correlation matrices ${\bf R}_{\rm tx}$ and ${\bf R}_{\rm rx}$ characterize the correlations among the sub-channels at the transmitter and at the receiver, respectively.

In this letter we consider the exponential correlation matrix in \cite{Lokya01a} 
with elements given as
\begin{equation} {\bf R}_{\rm rx}[j,i]=\begin{cases}  \rho_{\rm rx}^{j-i}, & i\leq j\\ (\rho_{\rm rx}^{i-j})^{\ast}, & i > j \end{cases},\quad i,j \in \{1,\ldots,n_{\rm R}\} \label{eq:Tx_correlation_mat} \end{equation}
where $\rho_{\rm rx}$ denotes the level of correlation at the Rx. The corresponding correlation matrix at the transmitter ${\bf R}_{\rm tx}$ can be obtained by replacing $\rho_{\rm rx}$ with $\rho_{\rm tx}$ in (\ref{eq:Tx_correlation_mat}). 

Let ${\bf A}_i$ be the $n_{\rm R} \times n_{\rm T}$ MIMO matrix with iid complex Gaussian random entries, i.e., ${\bf A}_i[r,t] \sim \mathcal{C}\mathcal{N}(0,\sigma_i^{2}(r,t))$, where $\sigma_i^{2}{(r,t)}$ denotes the average power associated with $i$-th path of the $(r,t)$ sub-channel. The spatially correlated channel coefficients in (\ref{eq:DD_io_hrt}) are then generated as  $h_i^{(r,t)}=\bar{\bf A}_i[r,t]$ where
\begin{equation}
    \bar{\bf A}_i={\bf C}_{\rm rx}\cdot{\bf A}_i\cdot{\bf C}_{\rm tx}^{\dag}
\end{equation}
and where the {\em correlation-shaping} matrices ${\bf C}_{\rm tx}$ and ${\bf C}_{\rm rx}$ are the lower-triangular matrices obtained by the Cholesky decomposition of ${\bf R}_{\rm tx}$ and ${\bf R}_{\rm rx}$, i.e., ${\bf R}_{\rm tx}={\bf C}_{\rm tx}\cdot{\bf C}_{\rm tx}^{\dag}$ and ${\bf R}_{\rm rx}={\bf C}_{\rm rx}\cdot{\bf C}_{\rm rx}^{\dag}$. 

    } 

\subsection{Input-output relations \label{sec:MIMO_IO}}

\subsubsection{Time domain}
 Let $\mathcal{L}^{(r,t)}=\{\ell_i^{(r,t)}\}$ for $i=1,\ldots,P^{(r,t)}$ be the set of distinct {\em normalized delays} in the $(r,t)$ sub-channel. Using (\ref{MIMO_DT_eqn_int}), the time domain input-output relation for one frame can be written as
\begin{align}
    {\bf r}^{(r)}[q]=\sum_{t=1}^{n_{\rm T}}\sum_{\ell \in \mathcal{L}^{(r, t)}} {g}^{(r,t)}[\ell,q]{\bf s}^{(t)}[q-\ell]+{\bf z}^{(r)}[q] \label{MIMO_tdio}
\end{align}
where $q=m+n(M+L_{\rm G})$ for $m=0,\ldots,M+L_{\rm G}-1$, $n=0,\ldots,N-1$ and ${\bf z}^{(r)}[q]$ is the AWGN noise in the $r$-th receive antenna.
Let ${\bf G}^{ (r,t)}$ be the time-domain channel matrix for the $(r,t)$ sub-channel, with entries 
\begin{equation}
    {\bf G}^{(r,t)}[m+nM,[m-\ell]_M+nM]={g}^{(r,t)}[\ell,m+n(M+L_{\rm G})] \label{MIMO_td}
\end{equation}
for $\ell \in \mathcal{L}^{(r,t)}$ and {\em zero} otherwise.
The modulo-$M$ operation $[\cdot]_M$ is due to the time-domain CP per block. In the case of ZP per block, ${\bf G}^{(r,t)}$ becomes a lower triangular matrix, i.e., ${\bf G}^{(r,t)}[q,[q-\ell]_{MN}]=0$ if $q<\ell$, \cite{Rake_DFE}. 

The time-domain input-output relation in (\ref{MIMO_tdio}) can be written in a simple matrix form as
\begin{align}
    {\bf r}^{(r)}=\sum_{t=1}^{n_{\rm T}}{\bf G}^{(r,t)}{\bf s}^{(t)}+{\bf z}^{(r)},\quad r=1,\ldots, n_{\rm R} \label{TD_MIMO_io}
\end{align}

\subsubsection{Delay-Doppler domain\label{DD_dom}}
 From (\ref{MIMO_vec}), the DD information symbols are related to the time domain samples:
\begin{align}
    {\bf s}^{(t)}={\bf P}\cdot({\bf I}_{ M}\otimes{\bf F}_N^{\dag})\cdot{\bf x}^{(t)},\text{  }
    {\bf r}^{(r)}={\bf P}\cdot({\bf I}_{ M}\otimes{\bf F}_N^{\dag})\cdot{\bf y}^{(r)}
    \label{MIMO_vector_txrx}
\end{align}
where ${\bf x}^{(t)}={\rm vec}\left(({\bf X}^{(t)})^{\rm T}\right)$, ${\bf y}^{(r)}={\rm vec}\left(({\bf Y}^{(r)})^{\rm T}\right)$ and ${\bf P}$ is the row-column interleaver permutation matrix given in \cite{Rake_DFE}.
Substituting (\ref{MIMO_vector_txrx}) in (\ref{TD_MIMO_io}), the corresponding DD domain input-output relation at the $r$-th receive antenna can be written as
\begin{align}
    {\bf y}^{(r)}=\sum_{t=1}^{n_{\rm T}}{\bf H}^{(r,t)}{\bf x}^{(t)}+{\bf w}^{(r)} \label{MIMO_IO}
\end{align}
\vspace{-2.2mm}
where 
\begin{align}
    {\bf H}^{(r,t)}&=({\bf I}_M\otimes{\bf F}_N)\cdot({\bf P}^{\rm T}\cdot{\bf G}^{(r,t)}\cdot{\bf P})\cdot({\bf I}_M\otimes{\bf F}_N^{\dag}) \\
    {\bf w}^{(r)}&=({\bf I}_M\otimes{\bf F}_N)\cdot({\bf P}^{\rm T}\cdot{\bf z}^{(r)})
\end{align}
To describe the proposed detection method (in Section \ref{sec:det}) we partition the $NM \times 1$ vectors ${\bf x}^{(t)}$ and ${\bf y}^{(r)}$ into $M$ {\em symbol-vectors} of length $N$ as
\begin{equation}
    {\bf x}^{(t)}=
  [{\bf x}_0^{(t){\rm T}},\ldots,
  {\bf x}_{M-1}^{(t){\rm T}}]^{\rm T} \\
,\quad {\bf y}^{(r)}=
  [{\bf y}_0^{(r){\rm T}},\ldots,
  {\bf y}_{M-1}^{(r){\rm T}}]^{\rm T} \\
\end{equation}
 Following the SISO-OTFS notations in \cite{Rake_DFE}, the input-output relation for MIMO-OTFS in (\ref{MIMO_IO}) can be written for each symbol-vector as
\begin{equation}
    {\bf y}_m^{(r)}=\sum_{t=1}^{n_{\rm T}}\sum_{\ell \in \mathcal{L}^{(r,t)}}{\bf K}_{m,\ell}^{(r,t)}\cdot{\bf x}_{[m-\ell]_M}^{(t)}+{\bf w}_m^{(r)}, \quad m=0,\ldots, M-1 \label{MIMO_OTFS_io}
\end{equation}
where ${\bf K}_{m,\ell}^{(r,t)} \in \mathbb{C}^{N \times N}$ is the $(r,t)$ sub-channel between the $m$-th received symbol vector of the $r$-th receive antenna and the $[m-\ell]_M$-th transmit symbol vector of the $t$-th transmit antenna, i.e., ${\bf K}_{m,\ell}^{(r,t)}$ is the $(m,[m-\ell]_M)-$th sub-matrix of ${\bf H}^{(r,t)}$ in (\ref{MIMO_IO}). It was shown in \cite{Rake_DFE} that for rectangular pulse shaping waveforms, ${\bf K}_{m,\ell}^{(r,t)}$ are circulant matrices.
\subsubsection{Delay-time domain}\label{DT:domain}
Here, we discuss the MIMO input-output relation in the DT domain, {\color{black} where detection can be performed with the least complexity} (see Section \ref{sec:det}). Vectors with a tilde denote the corresponding DT domain symbol-vectors and are related by the $N$-point DFTs as 
\begin{equation}
    \widetilde{\bf x}_m^{(t)}={\bf F}_N^{\dag}\cdot{\bf x}_m^{(t)},\quad\widetilde{\bf y}_m^{(r)}={\bf F}_N^{\dag}\cdot{\bf y}_m^{(r)}, \label{DT_vec}\vspace{-1 mm}
\end{equation}
 Since the DD domain sub-matrices ${\bf K}_{m,\ell}^{(r,t)}$ are circulant with ${\bf h}_{m,\ell}^{(r,t)} \in \mathbb{C}^{N \times 1}$ as their first column, (\ref{MIMO_OTFS_io}) can be written in form of element-wise multiplication in the corresponding Fourier transformed domain (i.e., the DT domain)  as
\begin{equation}
    \widetilde{\bf y}_m^{(r)}[n]=\sum_{t=1}^{n_{\rm T}}\sum_{\ell \in \mathcal{L}^{(r,t)}}\widetilde{\bf h}_{m,\ell}^{(r,t)}[n]\widetilde{\bf x}_{[m-\ell]_M}^{(t)}[n]+\widetilde{\bf w}_m^{(r)}[n] \label{MIMO_OTFS_io_DT}\vspace{-3 mm}
\end{equation}
for $n=0,\ldots,N-1$, where $\widetilde{\bf w}_m^{(r)}$ is the AWGN noise and  $\widetilde{\bf h}_{m,\ell}^{(r,t)}={\bf F}_N^{\dag}\cdot{\bf h}_{m,\ell}^{(r,t)}$ are the DT channel vectors, \cite{Rake_DFE}.


\section{MIMO-OTFS Detection}\label{sec:det}
This section proposes a low-complexity linear diversity-combining detector for MIMO-OTFS in the DT domain, based on the MRC principle. Consider the DT domain input-output relation in (\ref{MIMO_OTFS_io_DT}). For ease of illustration, we consider the number of distinct delay taps in all the sub-channels to be equal, i.e., $L=|\mathcal{L}^{(r,t)}|$  $\forall$ $r,t$. Then, due to multipath and spatial diversity, $L$ copies of each transmitted symbol-vector ${\bf x}_m^{(t)}$ arrive at each of the $n_{\rm R}$ receiver antennas along with multipath echoes of other symbol-vectors due to inter-delay and inter-antenna interference. The basic idea of the proposed detection method is to extract and combine the received signal components in all the diversity branches to improve the signal to interference plus noise ratio (SINR) of the desired signal in each iteration. To clearly view the desired signal and interference components in each branch for a given $\ell$ and $r$, the input-output relation in (\ref{MIMO_OTFS_io_DT}) is rewritten for $n=0,\ldots,N-1$  as
\begin{align}
    \widetilde{\bf y}_{m+\ell}^{(r)}[n]&=\sum_{t=1}^{n_{\rm T}}\sum_{\ell^{\prime}}\widetilde{\bf h}_{m+\ell,{\ell^{\prime}}}^{(r,t)}[n]\widetilde{\bf x}^{(t)}_{m+\ell-\ell^{\prime}}[n]+\widetilde{\bf w}_{m+\ell}^{(r)}[n] \nonumber \\
    &=\widetilde{\bf h}_{m+\ell,\ell}^{(r,t)}[n]\widetilde{\bf x}^{(t)}_{m}[n]+\widetilde{\bf v}_{m,\ell}^{(r,t)}[n]+\widetilde{\bf w}_{m+\ell}^{(r)}[n] 
    \label{MIMO_OTFS_io2} 
\end{align}
for $r=1,\ldots,n_{\rm R}$ and $ \ell \in \mathcal{L}^{(r,t)}$, 
where ${\bf w}_{m+\ell}^{(r)}$ is the AWGN noise vector, and the $N\times 1$ {\em interference vector}:
\begin{align}
\widetilde{\bf v}_{m,\ell}^{(r,t)}[n]=\underbrace{\sum_{\ell' \in \mathcal{L}^{(r,t)},\ell'\neq \ell}\widetilde{\bf h}_{m+\ell,\ell'}^{(r,t)}[n]\widetilde{\bf x}_{[m+\ell-\ell']_M}^{(t)}[n]}_\text{inter-delay interference} \nonumber \\+\underbrace{\sum_{t' \neq t}\sum_{\ell' \in \mathcal{L}^{(r,t')}}\widetilde{\bf h}_{m+\ell,\ell'}^{(r,t')}[n]{\bf x}_{[m+\ell-\ell']_M}^{(t')}[n]}_\text{inter-antenna interference} \label{MIMO_Interference}
\end{align}

Let $\widetilde{\bf b}_{m,l}^{(r,t)}[n]$ be the interference-cancelled component of $\widetilde{\bf x}_m^{(t)}$ received in the $\ell$-th delay branch of the $(r,t)$-th sub-channel: 
\begin{equation}
    \widetilde{\bf b}_{m,l}^{(r,t)}[n]=\widetilde{\bf y}_{m+\ell}^{(r)}[n]-\widetilde{\bf v}_{m,\ell}^{(r,t)}[n] \label{eq:b_ml}
\end{equation}
If the estimates of the transmitted time domain samples $\widetilde{\bf x}_m^{(t)}$ are available, then the interference $\widetilde{\bf v}_{m,\ell}^{(r,t)}$ can be computed from (\ref{MIMO_Interference}) and substituted in (\ref{eq:b_ml}) to cancel the inter-antenna and inter-delay interference. However, since the information symbols are unknown at the receiver, the interference is only partially cancelled. To improve the SINR, we maximal ratio combine the received copies of $\widetilde{\bf x}_m^{(t)}$ in each iteration as: 
\begin{equation}
    \hat{\widetilde{\bf x}}_m^{(t)\{i\}}[n]=\frac{\sum_{r=1}^{n_{\rm R}}\sum_{\ell \in \mathcal{L}}{\widetilde{\bf h}_{m+\ell,\ell}^{(r,t)\ast}[n]{\bf b}_{m,l}^{(r,t)\{i\}}[n]}}{\sum_{r=1}^{n_{\rm R}}\sum_{\ell \in \mathcal{L}}|\widetilde{\bf h}_{m+\ell,\ell}^{(r,t)\ast}[n]|^2} \label{eq:MRC_DT}
\end{equation}
where $\widetilde{\bf b}_{m,l}^{(r,t)\{i\}}$ is the interference-cancelled copy of $\widetilde{\bf x}_m^{(t)}$ in the $i$-th iteration computed using the current estimates $\hat{\widetilde{\bf x}}_m^{(t)\{i\}}$.
Let us define $\hat{\widetilde{\bf y}}_m^{(r){\{i\}}}$ to be the reconstructed received waveform from the current estimates of the symbol-vectors:
\begin{equation}
    \hat{\widetilde{\bf y}}_{m+\ell}^{(r){\{i\}}}[n]=\sum_{r=1}^{n_{\rm R}}\sum_{\ell \in \mathcal{L}^{(r,t)}}\widetilde{\bf h}_{m+\ell,\ell}^{(r,t)}[n]\cdot\hat{\widetilde{\bf x}}_{m}^{(t){\{i\}}}[n]
\end{equation}
From (\ref{MIMO_OTFS_io2}) and (\ref{MIMO_Interference}), the interference-cancelled component in (\ref{eq:b_ml}) of ${\bf x}_m^{(t)}$ in the $(i+1)$-th iteration can be written as 
\begin{equation}
    {\bf b}_{m,\ell}^{(r,t){\{i+1\}}}[n]=
   \widetilde{\bf h}_{m+\ell,\ell}^{(r,t)}[n]\cdot\widetilde{\bf x}_{m}^{(t){\{i\}}}[n]+
   \Delta\widetilde{\bf y}_{m+\ell}^{(r)}[n]
\label{b_ml_reduced}
\end{equation}
where ${\Delta\widetilde{\bf y}_{m+\ell}^{(r)}} =\widetilde{\bf y}_{m+\ell}^{(r)}[n]-\hat{\widetilde{\bf y}}_{m+\ell}^{(r){\{i\}}}[n] \in \mathbb{C}^{N \times 1}$ is the error in reconstructing the received DT waveform from current symbol-vector estimates. Substituting (\ref{b_ml_reduced}) in (\ref{eq:MRC_DT}), we get the MRC estimate of the DT samples for the next iteration 
\begin{align}
    \hat{\widetilde{\bf x}}_m^{(t){\{i+1\}}}[n]&=\hat{\widetilde{\bf x}}_m^{(t){\{i\}}}[n]+\nonumber \\&\widetilde{\bf d}_{m}^{(t)}[n]\left(\sum_{r=1}^{n_{\rm R}}\sum_{\ell \in \mathcal{L}^{(r,t)}}\widetilde{\bf h}_{m+\ell,\ell}^{(r,t)\ast}[n]\cdot\Delta\widetilde{\bf y}_{m+\ell}^{(r)}[n]\right) \label{MRC_DD_step1}
\end{align}
where
$  \widetilde{\bf d}_{m}^{(t)}[n]=\left(\sum_{r=1}^{n_{\rm R}}\sum_{\ell \in \mathcal{L}^{(r,t)}}|\widetilde{\bf h}_{m+\ell,\ell}^{(r,t)}[n]|^2\right)^{-1}
$.
For the first iteration, we can assume all $\hat{\widetilde{\bf x}}_m^{(t){\{0\}}}[n]=0$ for $n=0,\ldots,N-1$.
Then, as each sample $\hat{\widetilde{\bf x}}_m^{(t){\{i\}}}[n]$ is estimated, the $n_{\rm R}L$ reconstruction error samples $\Delta\widetilde{\bf y}_{m+\ell}^{(r)}[n]$ for $\ell \in \mathcal{L}^{(r,t)}$ and $r=1,\ldots,n_{\rm R}$ need to be updated:
\begin{align}
    \Delta\widetilde{\bf y}_{m+\ell}^{(r)}[n]\leftarrow \Delta\widetilde{\bf y}_{m+\ell}^{(r)}[n]-\widetilde{\bf h}_{m+\ell,\ell}^{(r,t)}[n]\Delta\widetilde{\bf x}_m^{(t){\{i+1\}}}[n] \label{MRC_DD_step2}
\end{align}
where $\Delta\widetilde{\bf x}_m^{(t){\{i+1\}}}[n]=\hat{\widetilde{\bf x}}_m^{(t){\{i+1\}}}[n]-\hat{\widetilde{\bf x}}_m^{(t){\{i\}}}[n]$.

 From (\ref{DT_vec}), the estimated DD information symbol-vectors at the end of $i$-th iteration is given by
 \begin{equation}
\hat{\bf x}_m^{(t){\{i+1\}}}=\hat{\bf x}_m^{(t){\{i\}}}+{\bf F}_N\cdot\Delta\widetilde{\bf x}_m^{(t){\{i+1\}}} \label{DT_to_DD}
\end{equation}
followed by a hard-decision $\mathcal{D}(\hat{\bf x}_m^{(t){\{i\}}})$ on the symbol-vectors to get the QAM estimates. 
To improve convergence speed,  we use a weighted average of the DD information symbol estimate with its hard decision as: ${\bf \hat{x}}_m^{(t){\{i\}}}\leftarrow (1-\delta){\bf \hat{x}}_m^{(t){\{i\}}}+\delta\mathcal{D}(\hat{\bf x}_m^{(t){\{i\}}})$. The iterations are stopped when the residual error does not decrease any more or when the maximum number of iteration is reached.

\subsection{Optimal combining weights for spatial correlation at Rx \label{sec:comb_weights}}

It is well known that MRC is optimal when the noise in the combining diversity branches is uncorrelated, \cite{MRC0,MRC2}. However in MIMO, due to spatial correlation, the interference in the diversity branches are correlated, resulting in performance degradation with MRC. In such scenarios, a whitening filter can be applied at the Rx to decorrelate the diversity branches. The MRC weights then need to be optimized to match the SINR of the whitened diversity branches. In this work, we consider only the correlation between the Rx antennas and assume that a whitening filter is applied at the Tx to mitigate the effect of the Tx correlation.

Assuming the Rx correlation matrix is known, we first group the combining weights corresponding to the $n_{\rm R}L$ diversity branches $\widetilde{\bf h}_{m+\ell,\ell}^{(r,t)}[n]$ for $\ell \in \mathcal{L}$ and $r \in \{1,\ldots,n_{\rm R}\}$ in (\ref{MRC_DD_step1}) into $L$ {\em antenna weight vectors} of length $n_{\rm R}$. The purpose of this grouping is to separate the diversity branches into correlated and uncorrelated branches. The combining step in (\ref{MRC_DD_step1}) can be written in terms of the $ n_{\rm R} \times 1$ {\em antenna weight vectors} for the $\ell$-th delay branch with entries $\widetilde{{\bf h}}_{m,n,\ell}^{(t)}[r]=\widetilde{\bf h}_{m+\ell,\ell}^{(r,t)}[n]$, as
 \begin{align}
    \Delta\widetilde{\bf x}_m^{(t){\{i+1\}}}[n]=\frac{\sum_{\ell}\widetilde{{\bf h}}_{m,n,\ell}^{(t)\dag}\Delta\widetilde{\bf y}_{m+\ell,n}}{\sum_{\ell}\widetilde{{\bf h}}_{m,n,\ell}^{(t)\dag}\widetilde{{\bf h}}_{m,n,\ell}^{(t)}} \label{MRC_DD_step1_vec}
\end{align}   
 where the $n_{\rm R} \times 1$ corresponding {\em antenna residual error vector} with components $\Delta\widetilde{\bf y}_{m+\ell,n}[r]=\Delta\widetilde{\bf y}_{m+\ell}^{(r)}[n]$ for $1 \leq r \leq n_{\rm R}$.

 We assume that the channel coefficients corresponding to different delay branches are uncorrelated, i.e., ${\rm E}[\widetilde{{\bf h}}_{m,n,\ell}^{(t)\dag}\widetilde{{\bf h}}_{m,n,\ell'}^{(t)}]=0$ for $\ell \neq \ell'$.  However, due to correlation between the receiver antennas, the components of each antenna weight vector are correlated, i.e., ${\bf R}_{\rm rx}={\rm E}[\widetilde{{\bf h}}_{m,n,\ell}^{(t)}\widetilde{{\bf h}}_{m,n,\ell}^{(t)\dag}]$ has non-zero off-diagonal elements. 

To alleviate performance loss, the antenna residual error vectors and the antenna weight vectors must be decorrelated before employing the MRC,\cite{MRC2}. To this end, consider the Cholesky decomposition of the covariance matrix ${\bf R}_{\rm rx}={\bf C}_{\rm rx}\cdot{\bf C}_{\rm rx}^{\dag}$, where ${\bf C}_{\rm rx}$ is the {\em correlation-shaping matrix}. Let ${\bf W}_{\rm rx}={\bf C}_{\rm rx}^{-1}$ be the {\em whitening matrix}. 
Then, the antenna weight and residual error vectors in (\ref{MRC_DD_step1_vec}) can be replaced with the corresponding whitened vectors as 
\begin{equation}
    \widetilde{{\bf h}}_{m,n,\ell}^{(t)}\leftarrow {\bf W}_{\rm rx} \cdot\widetilde{{\bf h}}_{m,n,\ell}^{(t)}, \quad \Delta\widetilde{\bf y}_{m}^{n}\leftarrow {\bf W}_{\rm rx} \cdot\Delta\widetilde{\bf y}_{m}^{n} \label{eq:decorr}
\end{equation}
The MRC combining operation in (\ref{MRC_DD_step1_vec}) is modified to include the decorrelation operation in (\ref{eq:decorr}) as
 
 \begin{align}
    \Delta\widetilde{\bf x}_m^{(t){\{i+1\}}}[n]&
    =\frac{\sum_{\ell}\widetilde{{\bf h}}_{m,n,\ell}^{(t)\dag}\cdot{\bf R}_{\rm rx}^{-1}\cdot\Delta\widetilde{\bf y}_{m+\ell,n}}{\sum_{\ell}\widetilde{{\bf h}}_{m,n,\ell}^{(t)\dag}\cdot{\bf R}_{\rm rx}^{-1}\cdot\widetilde{{\bf h}}_{m,n,\ell}^{(t)}}
    \label{MRC_DD_step1_vec_upd}
\end{align}

If we assume that the Rx has no prior knowledge of the correlation parameters, the correlation matrix can be estimated at the Rx for each frame. First the channel between all Tx-Rx pairs are estimated based on the single pilot method proposed in \cite{Choks_MIMO1}.  The correlation between the channel observed at the $r$-th and $r'$-th Rx antennas is computed from the estimated DT channel coefficients for  $r,r' \in \{1,\ldots n_{\rm R} \}$ as
\begin{equation}
    \hat{\bf R}_{\rm rx}[r,r']\!=\!{\rm E}[\hat{\widetilde{\bf h}}_{m,l}^{(r',t)\dag}\hat{\widetilde{\bf h}}_{m,l}^{(r,t)}] \!\propto \!\sum_{t=1}^{n_{\rm T}}\sum_{\ell \in \mathcal{L}}\sum_{m=0}^{M-1}\!\!\frac
    {\hat{\widetilde{\bf h}}_{m,\ell}^{(r',t)\dag}\hat{\widetilde{\bf h}}_{m,\ell}^{(r,t)}}
    {|\hat{\widetilde{\bf h}}_{m,\ell}^{(r',t)}||\hat{\widetilde{\bf h}}_{m,\ell}^{(r,t)}|} \label{eq:covar}
\end{equation}

{\bf Detection Complexity}: 
 We now discuss the complexity of the proposed detection method. The core steps of the MRC method without the whitening operation are (\ref{MRC_DD_step2}) and (\ref{MRC_DD_step1_vec}), and with whitening are (\ref{MRC_DD_step2}) and (\ref{MRC_DD_step1_vec_upd}). To distinguish both the methods, we will refer to the method using (\ref{MRC_DD_step1_vec}) as 'MRC' and the one with (\ref{MRC_DD_step1_vec_upd}) as 'MRCw'. The operation in (\ref{MRC_DD_step2}) requires $n_{\rm R}L$ complex multiplications (CM) per transmitted information symbol per iteration. The denominator of (\ref{MRC_DD_step1_vec}) and (\ref{MRC_DD_step1_vec_upd}) needs to calculated only once and requires $n_{\rm R}L$ CM and $(n_{\rm R}^2+n_{\rm R})L$ CM, respectively. The term $\widetilde{{\bf h}}_{m,n,\ell}^{(t)\dag}\cdot\hat{\bf R}_{\rm rx}^{-1}$ is computed only once and reused for all iterations. The estimation of ${\bf R}_{\rm rx}$ in (\ref{eq:covar}) and then ${\bf R}_{\rm rx}^{-1}$ requires $3NMn_{\rm T}n_{\rm R}L+O(n_{\rm R}^3)$ CMs. Then. the numerator in both (\ref{MRC_DD_step1_vec}) and (\ref{MRC_DD_step1_vec_upd}) requires only $n_{\rm R}L$ CM per transmitted information symbol per iteration. At the end of each iteration the hard decision estimates in (\ref{DT_to_DD}) requires $n_{\rm T}MN\log_2(N)$ CMs per iteration. 
 
 Assuming $S$ iterations are required, the overall number of CMs required for detecting all information symbols including the correlation matrix estimation operation is $n_{\rm T}NM [(4n_{\rm R}+n_{\rm R}^2)L+ S(3n_{\rm R}L+2\log_2 N+1)]+O(n_{\rm R}^3)$ CMs. This is significantly lower than the complexity of detection in MP ($O(n_{\rm T}n_{\rm R}^2NMSP^2Q)$) and linear minimum mean square error (LMMSE) ($O((n_{\rm T}n_{\rm R}NM)^3)$ detectors.
\section{Simulation Results and Discussion}
\begin{figure}
\centering
\hspace{-5mm}
{\includegraphics[trim=0 5 0 15,clip,height=2.1in,width=3.4in]{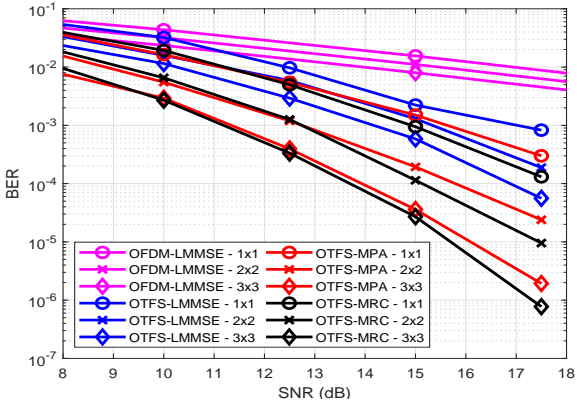}
 \vspace{-3mm}\caption{ 4-QAM MIMO-OTFS uncoded BER performance for MRC detector compared with LMMSE and MP detectors for a frame size of $N=M=32$ for different number of antennas.}
\label{BER_comp_MIMO2}}
\end{figure}

\begin{figure}
\centering
\hspace{-5mm}{\includegraphics[trim=0 5 0 15,clip,height=2.1in,width=3.4in]{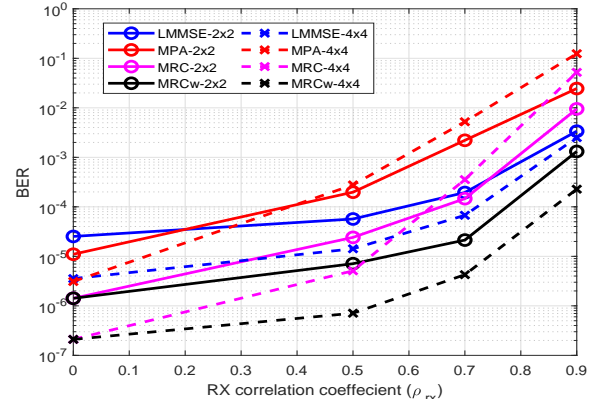}
 \vspace{-3mm}\caption{4-QAM  $2 \times 2$ and $4 \times 4$ MIMO-OTFS uncoded BER performance at a SNR of 20 dB for a frame size of $N=M=32$ for different Rx correlation levels $\rho_{\rm rx}$.}
\label{BER_comp_corr}}
\end{figure}

In this section, we present the uncoded BER performance of MIMO-OTFS\footnote{\color{black}We consider $N$ ZPs per frame, noting that when practical channel estimation is used, this results in the same overhead as OTFS with a single CP per frame since the ZPs are used anyway as part of the guard symbols around the pilot.} with the proposed detector and compare it with the MP and LMMSE detection methods. We generate OTFS frames of size $N=M=32$. The sub-carrier spacing $\Delta f$ is taken as $15\,$kHz, and the carrier frequency is set to $4\,$GHz. The number of paths in the channel $P^{(r,t)}$ is taken to be 5 with an uniform power delay profile with the set of delay taps $\mathcal{L}^{(r,t)}=\{0,\ldots,4\}$ and the Doppler shift for each path $\nu_{i}=\nu_{\rm max}\cos{\theta_i}$, with $\theta_i$ uniformly distributed over $(-\pi,\pi)$, where $\nu_{\rm max}$ is the maximum Doppler shift corresponding to a maximum UE speed of 500 km/hr. For BER plots, $10^{5}$ frames are sent for every point in the BER curve. In Figs.~\ref{BER_comp_MIMO2} and \ref{BER_comp_corr}, we assume perfect knowledge of the channel 
at the receiver.

In Fig.~\ref{BER_comp_MIMO2}, we present the 4-QAM BER performance of OTFS-MIMO with the proposed MRC detector and compare it with MP and LMMSE detectors. The maximum number of iterations is set to 20 for both MRC (with $\delta=0.125$) and MP. {\color{black}The MIMO-OFDM performance with LMMSE detector is plotted alongside to show the superior performance of OTFS in high mobility channels.} It can be observed that even though MP offers slightly better performance at very low SNR, MRC performs better at high SNR for the same number of iterations. We observed via simulations that both MP and MRC performance improve allowing for more iterations. 

\begin{figure}
\centering
\hspace{-5mm}{\includegraphics[trim=0 5 0 15,clip,height=2.1in,width=3.4in]{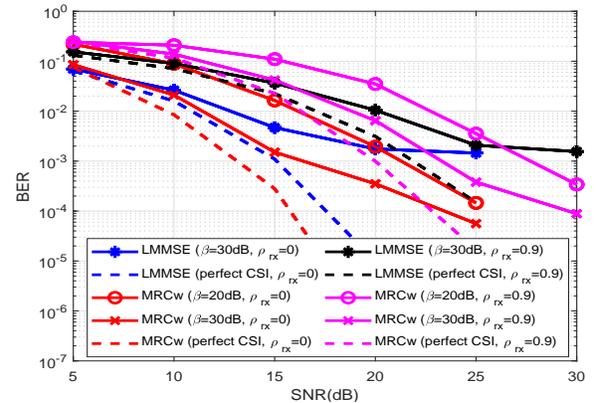}
 \vspace{-3mm}\caption{ 4-QAM  $2\times 2$ MIMO OTFS uncoded BER performance with MRC and LMMSE detector for a frame size of $N=M=32$ for excess pilot power ($\beta$) and Rx spatial correlation coefficient $\rho_{\rm rx}=0$ and $\rho_{\rm rx}=0.9$.}
\label{BER_comp_CSI}}
\end{figure}

{\color{black} Fig. \ref{BER_comp_corr} shows the $2 \times 2$ and $4 \times 4$ MIMO-OTFS performance of MRC compared with LMMSE and MPA detectors for different values of Rx spatial correlation ($\rho_{\rm rx}$). The curves labelled as `MRCw' denote the proposed detector with combining weights optimized according to the estimated Rx correlation matrix in (\ref{MRC_DD_step1_vec_upd}), while 'MRC' denotes the curves without the use of the whitening matrices for combining using (\ref{MRC_DD_step1_vec}). It can be observed that spatial correlation degrades the performance of all the detectors. However the proposed MRCw detector offers the best performance followed by the LMMSE detection. Both the MPA and MRC detection suffer some degradation in spatially correlated channels.

Fig.~\ref{BER_comp_CSI} presents the 4-QAM $2\times 2$ MIMO-OTFS BER performance for low ($\rho_{\rm rx}=0$) and high ($\rho_{\rm rx}=0.9$) correlation at the Rx. We consider practical channel estimation, where the channel coefficients are obtained using the single pilot method proposed in \cite{Choks_MIMO1,Ravi3}. The pilot symbol energy for each OTFS frame is given as $E_{\rm p}=\beta E_{\rm s}$, where $E_{\rm s}$ is the average symbol energy. The LMMSE detection performance is plotted alongside for comparison. The quality of the channel estimation depends on the pilot power as observed in this figure. It can be observed that the MRCw detector offers around 5dB gain   compared to LMMSE for the same excess pilot power $\beta=30dB$ for both low and high correlation at the Rx. For the perfect CSI case (dashed lines), it can be noted for both MRCw and LMMSE, that a spatial correlation of 0.9 causes a performance degradation of around 7 dB due to reduction in available space diversity as compared to the case with no correlation. In both cases MRCw gains 2dB over LMMSE at a much lower complexity.}
\vspace{-2mm}
\section{Conclusion}
In this paper, we proposed a low complexity detection method for MIMO-OTFS based on the MRC principle. The detection complexity was shown to be linear in number of information symbols and the number of receive and transmit antennas. We showed that the detector offers better error performance than MP and LMMSE detection methods with significantly lower complexity even with spatially correlated channels and practical channel estimation.

\end{document}